\documentclass[letter]{aa}
 \usepackage{txfonts}
 \usepackage{graphicx}
 \usepackage{amssymb}
 \usepackage{epsfig}
 \usepackage{natbib}
 \usepackage{tabularx}
 \usepackage{prettyref}
 \usepackage{textcomp}
 \usepackage{color}
 
\newcommand{\pref}{\prettyref}

 \newcommand{\ebv}{$E_{B-V}$}
 \newcommand{\lam}{$\lambda$}

\newcommand{\be}{\begin{equation}}
 \newcommand{\ee}{\end{equation}}
 
 \newcommand{\eg}{\emph{e.g.}}
 \newcommand{\kms}{\mbox{km\ \ensuremath{\rm{s}^{-1}}}}

\begin{document}
 
\title{Detection of diffuse interstellar bands in M31}
 
\author{M. A. Cordiner\inst{1}
 \and N. L. J. Cox\inst{2}
 \and C. Trundle\inst{1}
 \and C. J. Evans\inst{3}
 \and I. Hunter\inst{1}
 \and N. Przybilla\inst{4}
 \and F. Bresolin\inst{5}
 \and F. Salama\inst{6}
 }
 \authorrunning{Cordiner, Cox, Trundle, Evans, Hunter, Przybilla, Bresolin, \& Salama}

\offprints{Martin Cordiner, \email{m.cordiner@qub.ac.uk}}
 
\institute{Astrophysics Research Centre, School of Mathematics and Physics,
 Queen's University, Belfast, BT7 1NN, U.K.
 \and Herschel Science Centre, European Space Astronomy Centre, ESA, P.O.Box 78, E-28691 Villanueva de la Ca\~nada, Madrid, Spain.
 \and UK ATC, Royal Observatory Edinburgh, Blackford Hill, Edinburgh, EH9 3HJ, U.K.
 \and Dr. Remeis-Sternwarte Bamberg, Sternwartstr. 7, D-96049 Bamberg, Germany
 \and Institute for Astronomy of the University of Hawaii, 2680 Woodlawn Drive, 96822 Honolulu, HI, U.S.A.
 \and Space Science Division, NASA Ames Research Center, Mail Stop 245-6, Moffett Field, California 94035, U.S.A.}
 
\date{Received December 21st 2007 / Accepted January 23rd 2008}

\abstract{}{We investigate the diffuse interstellar band (DIB) spectrum in the interstellar medium of M31.
 }{%
 The DEIMOS spectrograph of the W. M. Keck observatory was used to make optical spectroscopic observations of two supergiant stars, MAG\,63885 and MAG\,70817, in the vicinity of the OB78 association in M31 where the metallicity is approximately equal to solar.
 }{%
 The \lam\lam5780, 5797, 6203, 6283 and 6613 DIBs are detected in both sightlines at velocities matching the M31 interstellar \ion{Na}{i} absorption. The spectra are classified and interstellar reddenings are derived for both stars. Diffuse interstellar band (DIB) equivalent widths and radial velocities are presented.
 }{%
 The spectrum of DIBs observed in M31 towards MAG\,63885 is found to be similar to that observed in the Milky Way. Towards MAG\,70817 the DIB equivalent widths per unit reddening are about three times the Galactic average. Compared to observations elsewhere in the Universe, relative to reddening the M31 ISM in the vicinity of OB78 is apparently a highly favourable environment for the formation of DIB carriers.}
 

\maketitle
 
\section{Introduction}
 
Today, more than 300 diffuse interstellar bands (DIBs) are known but the carriers have remained unidentified since their discovery almost 100 years ago. It is debated whether the DIB carriers arise from the dust, the gas, or the large-molecule component of the interstellar medium (see the review by \citealt{sarre06}). The substructure present in many of the DIB profiles indicates \citep{sarre95,ehren96} that they are caused by large gas-phase molecules. The interaction between gas-phase species (atoms and molecules), UV radiation and dust grains may be crucial in the formation of large ($\gtrsim50$ atom) molecules in space which are likely candidates for the carriers of the DIBs (see \emph{e.g.} \citealt{salama96,ruiter05}). 

In interstellar environments distinct from those found in the Milky Way (MW), previous research on the relationships between atoms, molecules, dust and DIBs has focused on the Large and Small Magellanic Clouds (\emph{e.g.} \citealt{ehren02,cox06,cox07,welty06}). The behaviour of the DIB carriers was analysed with respect to the higher gas-to-dust ratios, lower metallicities, lower $R_V$ and stronger interstellar radiation fields of these environments. Beyond the Magellanic Clouds, studies are sparse, confined to sightlines probed by sufficiently bright supernovae \citep[\eg][]{rich87,sollerman05} or background quasars (\emph{e.g.} \citealt{york06}).
 
M31 presents the opportunity to study the effects on the DIB carriers of the unique chemical and physical conditions found in this Local Group galaxy. This letter presents spectra of two supergiant stars in M31 and we report the first unambiguous detection of DIBs in M31. The observed DIB properties are discussed in relation to those in other galaxies and the interstellar conditions of M31.
 
\begin{table*}
 \centering
 \caption[Stars observed]{Observed M31 stars. Target numbers from \citet{magnier92}. Co-ordinates and photometry from \citet{massey06}.
 Derived stellar spectral classifications and radial velocities ($v_\ast$) are shown. M31 \ebv\ values were calculated using intrinsic colours from \citet{johnson66} and have been corrected for Galactic foreground reddening (see Sect.~\ref{sec:reddening}). }
 \label{tab:stars}
 \begin{tabular}{lcccccccc}
 \hline\hline
 Target&RA&DEC&$V$&$B-V$&Sp. Type&\ebv&$v_\ast$&S/N\\
 MAG\,\#&J2000&J2000&(mag)&(mag)&&(mag)&(\kms)&@5800~\AA\\\hline
 63885&00:40:30.47&+40:35:03.8&16.380$\pm$0.003&0.326$\pm$0.003&B9 Ia&$0.28\pm0.04$&$-544$&150\\
 70817&00:40:35.46&+40:36:44.6&18.282$\pm$0.003&0.431$\pm$0.003&F2 I&$0.12\pm0.04$&$-543$&80\\
 \hline
 \end{tabular}
 \end{table*}
 
\section{Observations}
 Seventy-two bright stars in and around the M31 OB\,78 association \citep{vandenBergh64} were observed in November 2003 using the Keck DEIMOS spectrograph. Two angles of the 1200G grating were used to cover the region from approximately 3500 to 9000 \AA\ with a resolving power $R=3300$. The total exposure time was 2.25 hours in the blue region and 1.5 hours in the red, during which the seeing was $0.5-0.8''$.\footnote{Initially the red and blue spectra were both reduced using the spec2d pipeline. However, the wavelength scale of the blue spectra ($\sim3500-6000$~\AA) was found to be erroneous so these were re-reduced manually in IRAF using the twodspec routines.} Wavelength calibration accuracy was confirmed using the Na D sky emission lines whose central wavelengths were found to be accurate to within $\pm5$ \kms. Reduced spectra were Doppler-corrected to the LSR frame. \pref{tab:stars} shows the co-ordinates and photometry of MAG\,63885 and MAG\,70817. These targets were selected for analysis in this letter due to their prominent DIBs. The rest of the DEIMOS data will be presented in a future article \citep{cord08}.

\section{Analysis and results}
 
\begin{figure}
 \centering
 \epsfig{file=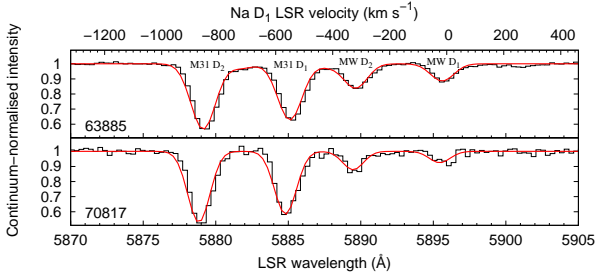, width=0.95\columnwidth}
 \caption{Observed continuum-normalised \ion{Na}{i} spectra (histograms) with best-fitting models overlaid (red curves). Velocity scale at the top is relative to the Na D$_1$ rest wavelength. The M31 and Galactic (MW) absorption components are labelled. }
 \label{fig:nai}
 \end{figure}

\subsection{Interstellar sodium D}\label{subsec:nai}
 
The observed Na D lines (see \pref{fig:nai}) show two clearly separated absorption components due to interstellar gas at velocities consistent with the Galactic and M31 ISM (around radial velocities of 0 and $-550$ \kms\ respectively). The spectra were modelled using \textsc{vapid} \citep{howarth02} and each absorption component was accurately reproduced with a Gaussian interstellar cloud model. The fits are plotted in \pref{fig:nai} and the mean radial velocities and equivalent widths (EWs) of the M31 D$_1$ absorption components are given in \pref{tab:dibs}.
 
\subsection{Diffuse interstellar bands}
 \label{sec:dibs}
 
We searched the spectra for all DIBs with central depths greater than 0.05 in Figure 6\footnote{A synthetic average of their observed DIB spectra (normalised to $E_{B-V}=1.28$).} of the DIB survey by \citet{jenn94}. Scaled to the reddening of our targets, this central depth limit corresponds to $\sim1\sigma$ of the spectral Poisson noise. The \lam\lam5780, 5797, 6203, 6283 and 6613 DIBs were detected at M31 velocities in the spectra of MAG\,63885 and MAG\,70817. \lam6196 was tentatively detected towards MAG\,70817. The \lam\lam4501, 4726, 5705, 5849, 6269, 6376, 6379, 6445, 6532, 6660, 6993 and 8026 DIBs were too weak to be detected. Telluric absorption line contamination prevented analysis of the \lam\lam6886, 6919, 7224 and 7334 DIBs.
 
\begin{figure}
 \centering
 \epsfig{file=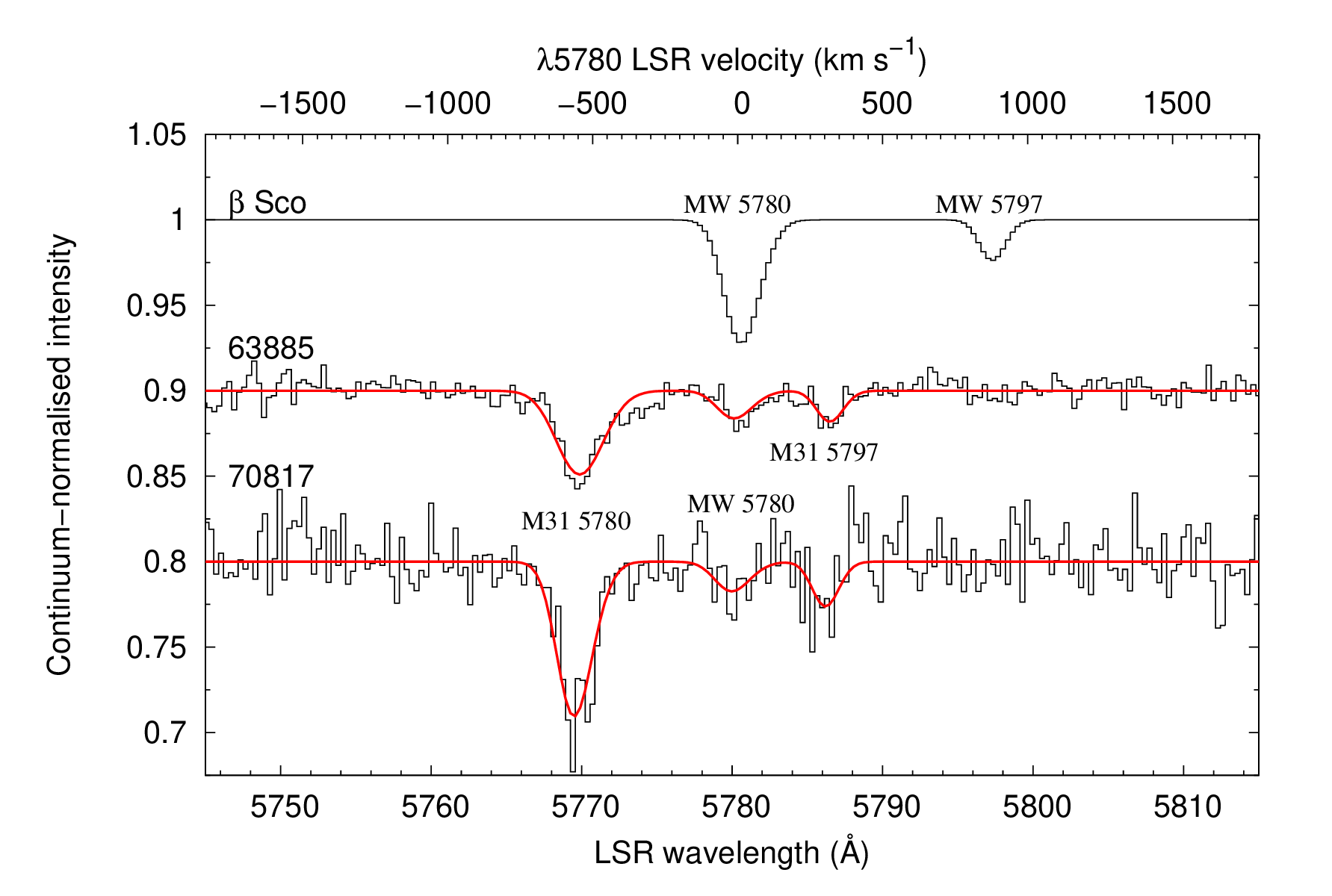, width=.9\columnwidth}
 \epsfig{file=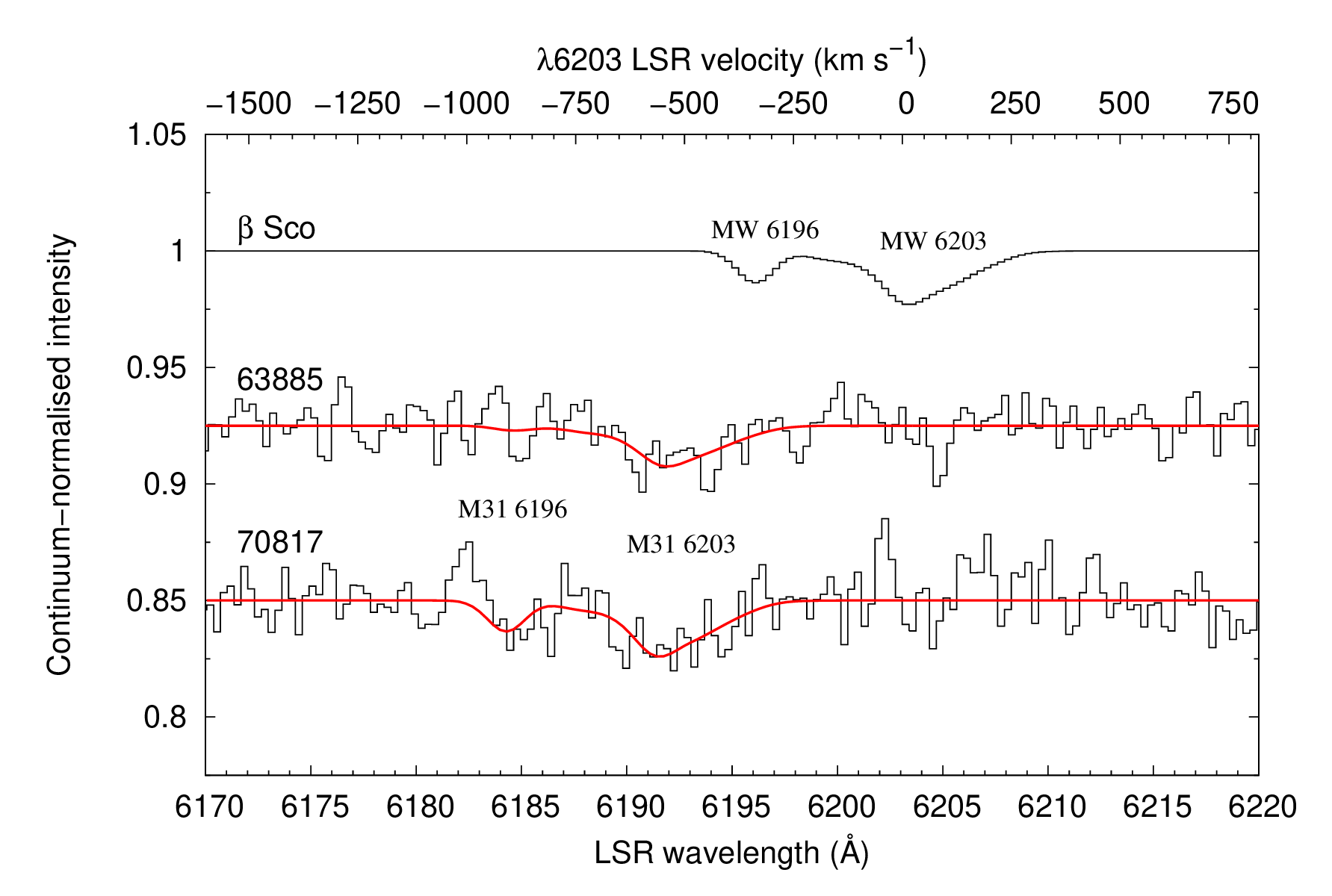, width=.9\columnwidth}
 \epsfig{file=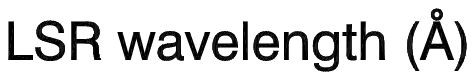,angle=90, width=.9\columnwidth}
 \epsfig{file=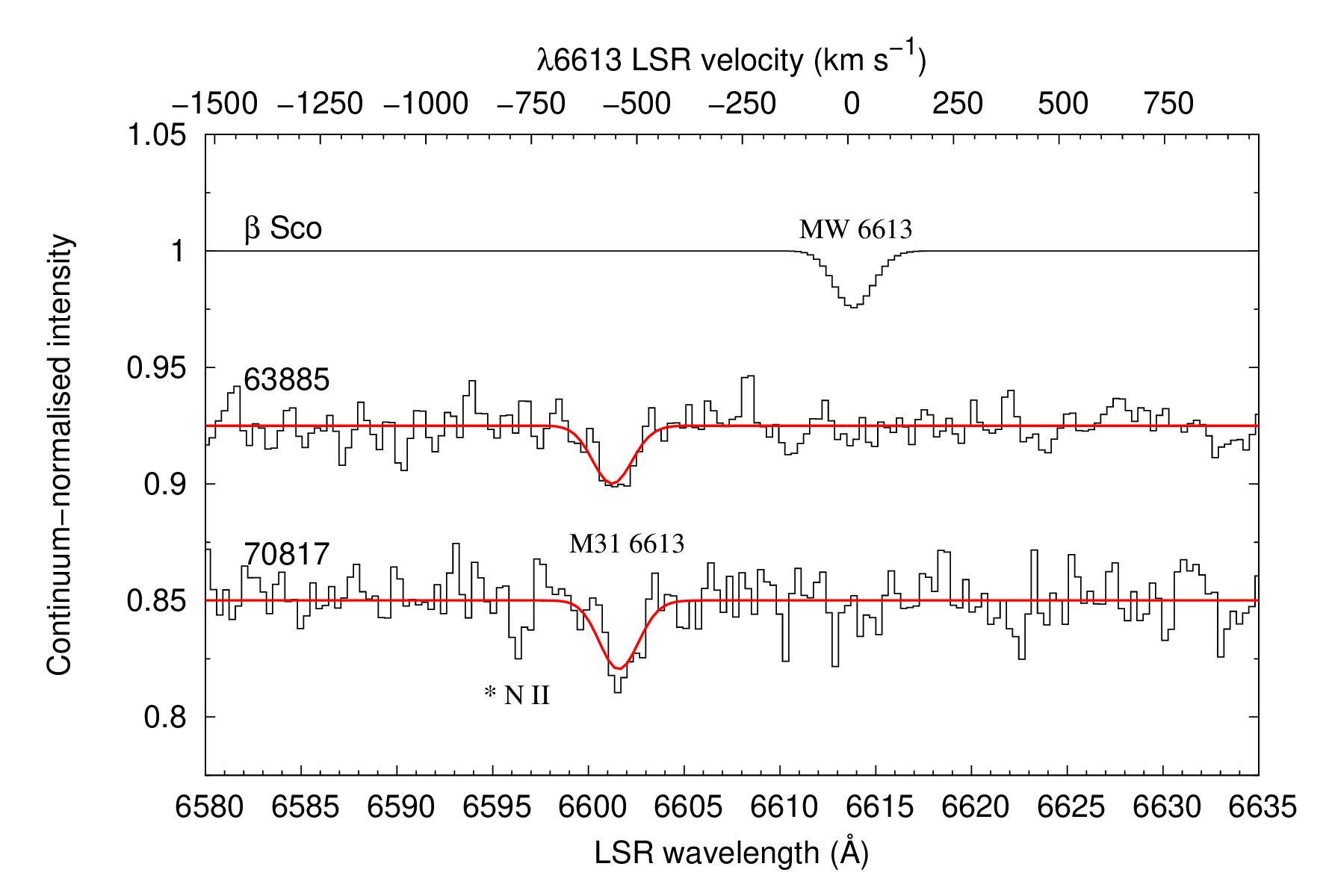, width=.9\columnwidth}
 \caption{Continuum-normalised spectra of the observed DIBs (black histograms). Galactic (MW) and M31 DIBs are labelled. Prominent stellar lines are marked with asterisks. Fitted model DIB profiles are overlaid (red curves). Velocity scales at the top are relative to the Galactic DIB rest wavelengths published by \citet{galaz00}. $\beta^1$ Sco Galactic reference DIB templates (plotted with arbitrary intensity scaling factors), are shown at the top of each figure, with additional telluric O$_2$-band spectrum (dashed line) for the \lam6283 region.}
 \label{fig:DIBs}
 \end{figure}

To simultaneously measure the DIB radial velocities and equivalent widths, Galactic DIB templates were fitted to the observed spectra using a non-linear least-squares algorithm. The Galactic templates were derived from high resolution, high S/N spectra of $\beta^1$\,Sco (see \citealt{cord06}). These were shifted to the interstellar rest frame and convolved with a Gaussian interstellar cloud model then with the instrumental spectral PSF and rebinned to the wavelength scale of the DEIMOS spectra. This technique minimises statistical errors in the measurement of the equivalent widths and velocities of weak DIBs but assumes that the intrinsic M31 DIB profiles and rest wavelengths closely match those of $\beta^1$\,Sco. If this is not the case, unknown systematic errors could occur, but we find no evidence to suggest that M31 DIB profiles differ at all from those of $\beta^1$\,Sco. Observed DIB spectra and fitted profiles are shown in \pref{fig:DIBs}.
 
For the \lam\lam5780, 6283 and 6196 DIBs the radial velocities and equivalent widths were allowed to vary in the fits. A telluric absorption component (derived from a high-resolution spectrum of the unreddened fast-rotator $\alpha$ Gru \citep[see][]{cord06}, degraded to $R=3300$), was also included for \lam6283 to account for the prominent band-head of atmospheric O$_2$ between 6278 and 6286 \AA.\footnote{Our DEIMOS observations contained no telluric standards by which to perform a conventional telluric division.} For the \lam\lam5797, 6196 and 6203 DIBs only the equivalent widths were allowed to vary while the DIB radial velocities were fixed at the M31 \ion{Na}{i} radial velocities. In all cases, there is excellent agreement between the fitted and observed DIB profiles. DIB radial velocities and EWs are shown in \pref{tab:dibs}. The radial velocities of all measured DIBs closely match the \ion{Na}{i} radial velocities, consistent with previous high-resolution studies of the Galactic and extragalactic ISM (such as those by \citealt{sollerman05} and \citealt{cox06, cox07}).
 
Statistical equivalent width and DIB velocity error estimates were derived by replicating the spectra 100 times with randomly generated Poisson noise added to each. The DIB model parameters were refitted for each replication and individual parameter errors were obtained from the $\pm68$th percentiles of the resulting parameter ranges.
 
We also found evidence for a broad absorption feature towards MAG\,63885 centred near 4430 \AA, identified as a blend of M31 and Milky Way \lam4428 DIBs.  This feature was modelled as a sum of two Lorentzian components with FWHM 17.25 \AA\ \citep[see][]{snow02} at $v=0$ and $v=-550$ \kms. The best-fitting model yields 1057~m\AA\ for the M31 component and 236~m\AA\ for the MW foreground component. However, this result is uncertain due to stellar line contamination and the relatively poor signal-to-noise in this spectral region.

\begin{table}
 \centering
 \caption{Summary of best-fitting DIB equivalent widths (EW\,/m\AA) and LSR radial velocities ($v$\,/\kms) for MAG\,63885 and MAG\,70817. $f$ denotes values held fixed during fitting. Statistical (68th percentile) DIB velocity errors are less than about $\pm10$ \kms. The M31 \ion{Na}{i} radial velocities and D$_1$ line EWs are also shown, including an estimated contribution due to stellar photospheric \ion{Na}{i} in square brackets. EWs for $\beta^1$\,Sco ($E_{B-V}=0.22$) and Galactic mean EW/\ebv\ data (normalised to $E_{B-V}=0.14$) are also given (data from \citealt{herbig93}, \citealt{thorb03}, \citealt{megier05} and \citealt{cord06}).
 }
 \label{tab:dibs}
 \resizebox{0.85\columnwidth}{!}{
 \begin{tabular}{lllllll}
 \hline\hline
 &\multicolumn{2}{c}{MAG\,63885}&\multicolumn{2}{c}{MAG\,70817}&$\beta^1$ Sco&MW Avg.\\
 \cline{2-3}\cline{4-5}
 &EW&$v$&EW&$v$&EW&EW\\
 \hline
 5780&$190^{+26}_{-16}$&$-539$&$271^{+17}_{-26}$&$-559$&142&75\\
 5797&$44^{+8}_{-5}$&$-550^f$&$58^{+32}_{-16}$&$-565^f$&17&22\\
 6196&$3^{+9}_{-3}$&$-550^f$&$29^{+19}_{-9}$&$-565^f$&12&10\\
 6203&$94^{+21}_{-16}$&$-550^f$&$121^{+15}_{-13}$&$-565^f$&57&23\\
 6283&$260^{+11}_{-11}$&$-575$&$541^{+21}_{-21}$&$-558$&330&159\\
 \vspace{0.15cm}
 6613&$63^{+7}_{-8}$&$-564$&$74^{+11}_{-12}$&$-548$&43&31\\
 Na\,D$_1$&736\,\scriptsize{[70]}&$-550$&846\,\scriptsize{[300]}&$-565$&147\\
 \hline
 \end{tabular}
 }
 \end{table}

\subsection{Stellar spectral types}\label{sec:stellartypes}
 Spectral types and stellar radial velocities for the target stars are presented in \pref{tab:stars}. Analysis of luminous B-type supergiants in M31 finds chemical abundances comparable to those in the solar neighbourhood \citep[\eg][]{trundle02}, so Galactic standards are appropriate. The spectra were classified with reference to the Galactic standards from \citet{eh03}, with luminosity classes assigned on the basis of the
 H$\gamma$ equivalent widths \citep{eh04}.
 
Stellar radial velocities are the mean results from manual measurements of the cores of the H$\alpha$ Balmer line and the Paschen lines. Statistical errors on the measured radial velocities are less than $\pm5$ \kms.
 
We searched Galactic stellar spectra of the same spectral types as our targets for the presence of lines overlapping the detected DIBs. No significant contamination of the \lam\lam6196, 6203, 6283 or 6613 DIBs is expected. The \lam\lam5780 and 5797 may suffer contamination of up to about 5 and 2~m\AA\ respectively as a result of overlapping lines of \ion{Fe}{i}.
 
\subsection{Foreground gas and dust}\label{sec:reddening}
 
LAB \ion{H}{i} data (\citealt{kalberla05}) for the nearest survey point in the direction of our sightlines shows N(\ion{H}{i}) = $5.91 \times 10^{20}$ cm$^{-2}$ over the velocity range of Galactic gas (from $-300$ to $100$~\kms). Equation 7 of \citet{burst78} is used to calculate the foreground reddening for lines of sight at latitudes away from the Galactic plane which yields $E_{B-V}=0.06$~mag towards our targets, identical to the value given by the foreground dust map of \citet{schleg98}.
 
The Galactic foreground \ion{Na}{i} column densities measured towards MAG\,63885 and MAG\,70817 are $3.7\times10^{12}$ and $1.9\times10^{12}$ cm$^{-2}$ respectively, which correspond to reddenings of 0.12 and 0.08~mag \citep{hobbs74}. However, there may be contamination of the Galactic \ion{Na}{i} profiles near $v=0$ due to sky-line subtraction residuals. Using a foreground reddening of 0.06, corrected M31 \ebv\ values are given in \pref{tab:stars}.

\begin{figure}
 \centering
 \epsfig{file=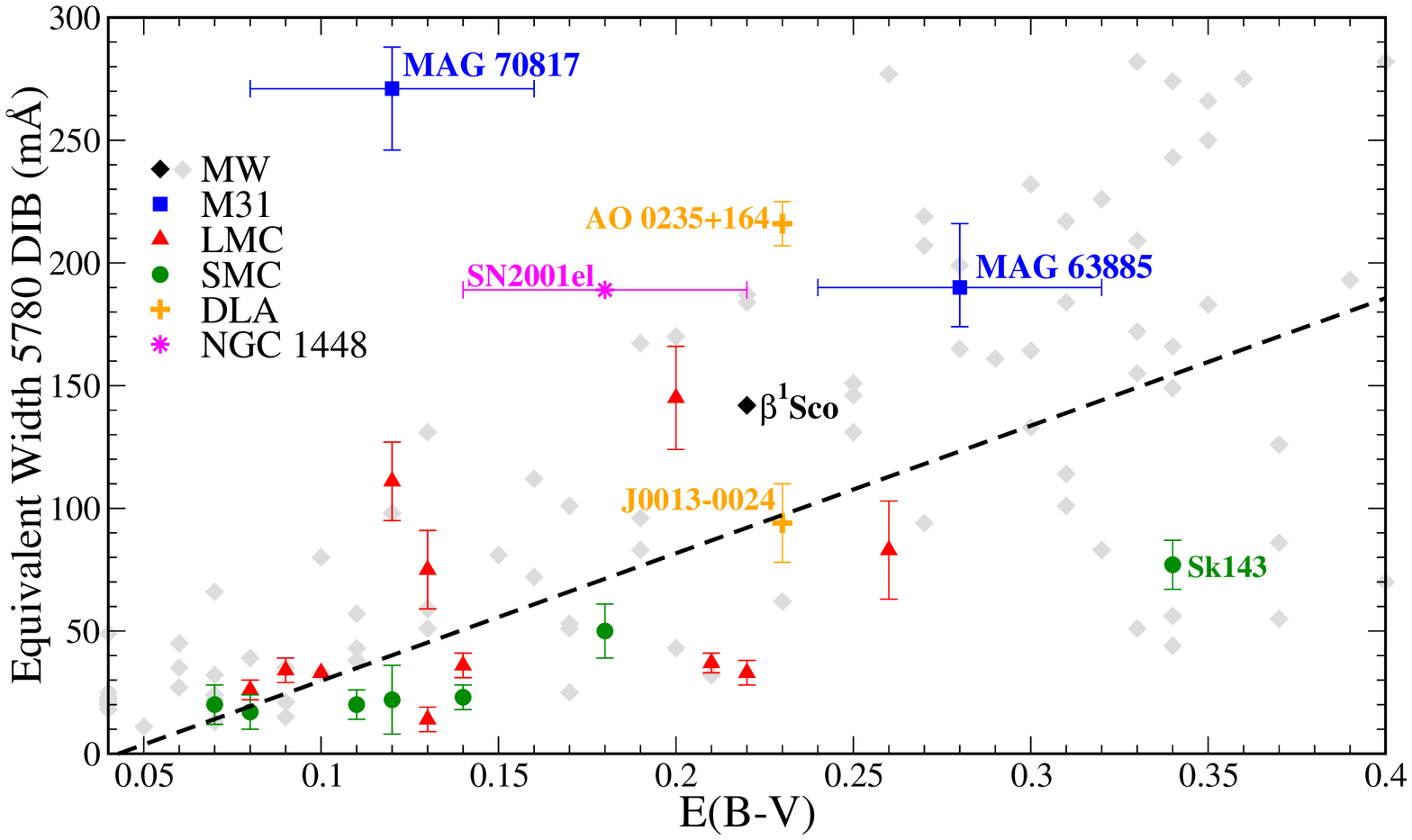, angle=-90,width=0.9\columnwidth}
 \caption{\lam5780 equivalent widths \emph{vs.} reddening for different galaxies. Data for the LMC and SMC \citep{welty06}, DLAs \citep{york06,ellison08} and NGC1448 \citep{sollerman05} are shown. The dashed line shows a linear fit to Galactic data (from \citealt{herbig93}, \citealt{thorb03}, \citealt{megier05} and \citealt{cord06}).}
 \label{fig:DIBCorrelationEbv}
 \end{figure}

\section{Discussion}
DIB equivalent widths in the Milky Way correlate with \ebv\ \citep{herbig95}. Given the signal-to-noise and the limited reddening of our target spectra, only the strongest known Galactic DIBs were detected in M31. Other M31 DIBs (including those listed in \pref{sec:dibs}) are expected in our spectra around or below the limit of detectability.
 
The DIB equivalent widths observed in M31 toward MAG\,63885 are consistent with those typically observed in Galactic sightlines with the same reddening, as highlighted by comparison with the similarly-reddened $\beta^1$\,Sco (see \pref{tab:dibs}). Towards MAG\,70817 the observed DIBs are stronger per unit \ebv\ by a factor of about $2-5$ compared with Galactic-mean data. \pref{fig:DIBCorrelationEbv} shows that the \lam5780 EW/\ebv\ for MAG\,63885 is among the larger values observed elsewhere in the Universe and for MAG\,70817 \lam5780 EW/\ebv\ is significantly greater than any other sightline plotted.
 
Detailed information is sparse, but M31appears be similar to the Milky Way in terms of the average interstellar gas-to-dust ratio (\citealt{nedia00a}, \citealt{bresolin02}) and the metallicity in the vicinity of our targets (\citealt{trundle02}). There is, however, evidence for a difference in the properties of the interstellar dust grains and the interstellar UV radiation field. A different star formation history and low rate of star formation (1/10th of MW; \citealt{walter94}) has been measured in M31, and the surface radiation flux was found to be poor in UV (\citealt{cesarsky98,pagani99}). The fact that strong DIBs are observed under these conditions may be in contradiction to the hypothesis that UV radiation is required for the production of the carriers \citep[see][for example]{herbig95,kendal02}. However, MAG\,63885 and MAG\,70817 are near to the OB78 association where the abundance of early-type stars may result in a strong interstellar UV radiation field.
 
The M31 2175 \AA\ UV extinction bump has been found to be weak and narrow \citep{bianchi96,hutch92} and a peculiar extinction law was observed by \citet{massey95} in the anomalously low average colour-excess ratio $E_{U-B}$/\ebv. Accepting current theories of dust grain extinction (see review by \citealt{draine03}), the evidence is consistent with a different distribution of dust grain sizes and compositions compared to the Milky Way average. In particular, there may be a lack of small graphitic dust grains (see also \citealt{xu94}). In that case, the observation of strong DIBs implies that the carriers are not closely associated with the small grains believed to be responsible for shape of the UV extinction curve.
 
\section{Conclusion}
 
The \lam\lam5780, 5797, 6283, 6203 and 6613 DIBs were detected towards MAG\,63885 and MAG\,70817 at velocities corresponding (within $25$~\kms) to the mean M31 interstellar \ion{Na}{i} absorption. \lam6196 and \lam4430 were also tentatively detected.
 
The M31 DIB spectrum towards MAG\,63885 is consistent with that observed in the Galaxy. Towards MAG\,70817 the DIBs are about a factor of three stronger per unit reddening than the Galactic average. The high DIB strengths might be related to differences in the M31 interstellar UV extinction curve and radiation field compared to the Galaxy. Further studies will be required to determine if the two sightlines examined here are representative of the general M31 DIB behaviour.
 
A planned instrument for the Hubble Space Telescope, the `Cosmic Origins Spectrograph' will be able to provide essential information on the UV properties of M31 sightlines to further examine these possibilities. However, the connection between DIBs in M31 and the atomic and molecular content (\eg\ Na, K, Ca, CH$^+$, CH, CN and C$_2$) of the diffuse medium can only be properly addressed with high resolution, high signal-to-noise optical spectroscopic observations.

\begin{acknowledgements}
 The W. M. Keck Observatory is operated as a scientific partnership among the California Institute of Technology, the University of California, and NASA. The observatory was made possible by the generous financial support of the W. M. Keck Foundation.
 The analysis pipeline used to reduce the Keck/DEIMOS data was developed at UC Berkeley with support from NSF grant AST-0071048.
 This research has made use of NASA's Astrophysics Data System.
 MAC and NLJC acknowledge support from the Faculty of the European Space Astronomy Centre (ESAC). MAC thanks QUB for financial support.
 
\end{acknowledgements}
 
\scriptsize
 \bibliographystyle{aa}
 \bibliography{M31refs}
 
\end{document}